\documentclass[12pt]{iopart}
\usepackage{amssymb}

\usepackage{epsfig}
\usepackage{rotating}
\usepackage{portland}
\usepackage{lscape}
\usepackage{sw20lart}
\usepackage[dvips]{tcigraph}

\input{tcilatex}
\begin{document}

\rotdriver{dvips}

\title{Light deflection in Weyl gravity: critical distances for photon
paths.}

\author{Sophie Pireaux
\footnote{%
Previously working in \newline
Unit\'{e} de Physique Th\'{e}orique et Math\'{e}matique (FYMA),\newline
Universit\'{e} catholique de Louvain (UCL), BELGIUM.}}

\address{%
UMR 5562, Dynamique Terrestre et Plan\'{e}taire (DTP), B105%
\newline
Observatoire Midi-Pyr\'{e}n\'{e}es,\newline
14 Avenue Edouard Belin,\newline
31400 Toulouse,\newline
FRANCE}

\ead{%
sophie.pireaux@cnes.fr}

\begin{abstract}The Weyl gravity appears to be a very peculiar
theory. The contribution of the Weyl linear parameter to the effective
geodesic potential is opposite for massive and nonmassive geodesics.
However, photon geodesics do not depend on the unknown conformal factor,
unlike massive geodesics. Hence light deflection offers an interesting test
of the Weyl theory.
In order to investigate light deflection in the setting of Weyl gravity, we
first distinguish between a weak field and a strong field approximation.
Indeed, the Weyl gravity does not turn off asymptotically and becomes even
stronger at larger distances.
We then take full advantage of the conformal invariance of
the photon effective potential to provide the key radial distances in Weyl
gravity. According to those, we analyze the weak and strong field regime for
light deflection. We further show some amazing features of the Weyl theory
in the strong regime.
\end{abstract}

\submitto{\CQG}

\pagebreak

\section{Introduction}

\qquad General Relativity, this monumental theory elaborated by Albert
Einstein, is a pure tensor theory which simply reduces to Newton's theory in
the weak field limit. It agrees, so far, with all the observations made in
our solar system and has to be taken into account in modern technologies
like the Global Positioning System (GPS) or when computing orbits.
Nevertheless, General Relativity cannot be the final theory of gravitation.
Indeed, solely from a theoretical point of view, several questions are
pending. First of all, the minimal choice of the Hilbert-Einstein action is
not based on any fundamental principle. Additionally, Einstein's theory of
gravitation cannot be properly described by quantum field perturbation
theory; and this makes it impossible to unify it with other fundamental
interactions. It is also not invariant under conformal transformations.
However, if we wish to achieve the unification of gravitation with particle
physics, in which conformal invariance plays a crucial role, we might
consider a theory of gravitation that incorporates this property.
Furthermore, nothing guarantees that the Newtonian potential is valid on
very short, or very long distances...\newline
From the point of view of experiments, even though Einstein's theory passed
so far all the relevant solar system tests (light bending, time delay,
perihelion shift of Mercury) within the experimental errors $\cite
{Will 2001 summary of tests}$, we recall that General Relativity cannot
reproduce the flat velocity distributions in the vicinity of galaxies
without using copious amounts of dark matter. The Newtonian potential would
indeed predict a decreasing distribution. We are thus confronted with the
following dilemma: either we assume the existence of a huge amount of dark
matter, or we modify the potential for galactic distances. This second
solution would immediately invalidate General Relativity with a null
cosmological constant. Note that the solution to the ``dark matter dilemma''
could also be a combination of the two solutions mentioned here above, thus
requiring a more acceptable amount of dark matter.\newline
Hence, an alternative theory of gravitation is highly desirable. Regarding
this demand, and among many other candidates, Weyl theory is an interesting
prototype. Not only does it present an interesting conformal invariance
property, but it also contains an additional linear contribution to the
Newtonian potential. This latter feature is encoded in the key parameter of
the theory, $\gamma _{W}$, and General Relativity is recovered for $\gamma
_{W}=0$.

Like any alternative theory of gravitation, Weyl theory needs to be tested.
But if as for various alternative theories, light deflection is an
interesting probe $\cite{Pireaux 2002 thesis}$, it proves to be even more
relevant when Weyl gravity is tackled. Indeed, similarly to other theories,
the sole Weyl gravitational action (no prescription for the matter action
needed) can be used to derive photon paths. It allows to deduce important
constraints on the theory free parameter. The crucial advantage of light
deflection in Weyl theory resides in the fact that such a study is free from
considerations on the unknown Weyl conformal factor. Whereas this factor, to
be provided by some symmetry breaking mechanism, must be specified to
analyze predictions involving the motion of massive particles, like galactic
rotation curves.

In the approach developed in the present article, we show how some critical
radii in Weyl theory structure the space-time with respect to photon paths.
However, we still leave open the question of the sign of the parameter $%
\gamma _{W}$. Indeed, we do not make any assumption on the conformal scale
factor, unlike in the approach used by Mannheim and Kazanas, who deduced a
positive value for $\gamma _{W}$ by fitting galactic rotation curves.
Nevertheless, light deflection provides us with some useful hints on the
physics implied by one or the other sign.\newline
We give here the orders of magnitude for the critical distances as implied
by the Mannheim-Kazanas parametrization ($\gamma _{W}>0$). In another
article \emph{``Light deflection in Weyl gravity: \textit{constraints on the
linear parameter}''} $\cite{Pireaux 2003 constraints on Weyl parameter}$,
we provide some less stringent constraints on $\gamma _{W}$ than Mannheim
and Kazanass, but unbiased by the scale factor for our constraints are based
solely on light deflection experiments. The orders of magnitude of critical
radii, weak and strong field limits can be recalculated accordingly.

\section{The Weyl theory}

\subsection{The gravitational action}

\qquad The Weyl theory (W) $\cite{Mannheim 1994 open questions},\,\cite
{Mannheim 1997 Global Gravity},\cite{Mannheim 1989 Exact Solution},\,\cite
{Mannheim 1991 Equations of motion}$ is a purely tensorial theory with a
metric coupling of gravitation to matter like in General Relativity (GR),
but with higher order derivatives of the metric in the dynamical sector. A
particularity of this theory is to be conformally invariant (under
transformations $\ g_{\mu \nu }\mapsto \chi ^{2}(x)\ g_{\mu \nu }$ with $%
\chi ^{2}(x)$ a finite, positive, nonvanishing, continuous real function),
which makes it attractive since renormalizable perturbatively $\cite
{Fradkin 1982 Weyl renormalizable},\,\cite{Stelle 1977 renormalization}$
and asymptotically free $\cite{Fradkin 1981 weyl asymp free},\,\cite{Julve
1978 weyl asymp free}$ in a way similar to the theory of strong
interactions.\newline
The Weyl gravitational action given by 
\begin{equation}
I_{W\ gravitation}\left. 
\begin{array}[t]{l}
=\int dx^{4}\sqrt{-g}\ W^{\mu \nu \rho \sigma }\ W_{\mu \nu \rho \sigma
}\smallskip \\ 
=\int dx^{4}\sqrt{-g}\ \left\{ R^{\mu \nu \rho \sigma }\ R_{\mu \nu \rho
\sigma }-2\ R^{\mu \nu }\ R_{\mu \nu }+\frac{1}{3}\ R^{2}\right\}
\end{array}
\right.  \label{WEYL_action}
\end{equation}
where $g$ it the determinant of the metric $g_{\mu \nu}$; $W^{\mu \nu \rho \sigma }$, 
$R^{\mu \nu \rho \sigma }$, $R^{\mu \nu }$ are the Weyl, Riemann and Ricci tensors 
respectively; and $R$, is the scalar of curvature associated with the metric. 
The latter action is conformally invariant since 
\[
W_{\mu \nu \rho \sigma }\mapsto \chi ^{2}(x)\ W_{\mu \nu \rho \sigma
}\,.\medskip 
\]

The geodesics for massive particles are obviously not conformally invariant,
but the way to implement this symmetry breaking is still obscure. However,
light geodesics are conformally invariant, see Subsection \ref
{light_versus_massive}, so that any gravitational test based on the
trajectories of photons (like the light deflection and the radar echo delay)
should not carry this ambiguity.\newline
In vacuum, the variational principle applied to action (\ref{WEYL_action})
with respect to the metric leads to the Bach equations: 
\begin{equation}
B^{\mu \nu }\equiv R_{\alpha \beta }\ W^{\mu \alpha \nu \beta }+2\ W^{\alpha
\mu \beta \nu }\ _{\mid \alpha \mid \beta }=0\,
\label{Bach_equation_Weyl}
\end{equation}
where $_{\mid \alpha}$ is the covariant derivative.
The vacuum solutions of GR ($R_{\alpha \beta }=0$) are also vacuum solutions
of the Weyl theory thanks to the well-known Bianchi identities. Thus, the
Schwarzschild metric is a particular solution of the spherically symmetric
Bach equations.\smallskip

The generalized Birkhoff theorem proves that there exists a three parameter
family ($\beta _{W}$, $\gamma _{W}$, $k_{W}$, all constants) of static and
spherically symmetric solutions to those Bach equations. Any of these
solutions can be recast, thanks to conformal and coordinate transformations
into a canonical line element $\cite{Mannheim 1989 Exact Solution}$, for a
efficient derivation using Killing vectors see $\cite{Pireaux 1997 memoire}$.
It provides the general solution to the Bach equations, 
\begin{eqnarray}
\fl ds^{2}=g_{\mu \nu }\ dx^{\mu }\ dx^{\nu }  \nonumber \\
\lo=\chi ^{2}(r)  \nonumber \\
\cdot \left\{ \left[ 1+\frac{2\,V_{W}}{c^{2}}\right] c^{2}dt^{2}-\left[ 1+%
\frac{2\,V_{W}}{c^{2}}\right] ^{-1}dr^{2}-r^{2}\left( d\theta ^{2}+\sin
^{2}\theta \,d\varphi ^{2}\right) \right\}  \label{metric_Schwarzschild_Weyl}
\end{eqnarray}
where 
\begin{equation}
\fl V_{W}(r)=\;-\frac{\beta _{W}}{2}\ \frac{\left( 2-3\beta _{W}\ \gamma
_{W}\right) }{r}c^{2}\;-\frac{3}{2}\ \beta _{W}\ \gamma _{W}\ c^{2}\;+\frac{%
\gamma _{W}}{2}r\ c^{2}\;-\frac{k_{W}}{2}r^{2}\ c^{2}\,\text{.}\smallskip
\label{effective_potential_WEYL}
\end{equation}
The Weyl metric contains, as expected, the Schwarzschild line element as a
particular case ($\gamma _{W}=k_{W}=0$ with $\beta _{W}(M)=\frac{G_{N}M}{%
c^{2}}$, where $G_{N}$ is the Newtonian constant, and $M$, the gravitational
mass). The conformal factor $\chi ^{2}(r)$ is arbitrary unless a conformal
symmetry breaking mechanism is specified.

\subsection{The gravitational potential}

\qquad The gravitational potential corresponding to the Weyl static and
spherically symmetric metric is not Newtonian anymore. If choosing\ $\chi
^{2}(r)\equiv 1$ (or choosing a constant $\chi ^{2}$, in which case $r$\ and 
$t$\ are rescaled by a constant factor), this is the so-called Weyl
gravitational potential (\ref{effective_potential_WEYL}). It presents a
constant ($\beta _{W}\ \gamma _{W}$)-term in addition to the traditional
Newtonian potential (1st term), so that the solution cannot, with any
conformal transformation, be brought to a Minkowskian space, even
asymptotically. It contains also a linear term that might dominate at
galactic distance scales; as well as a quadratic term that should become
significant only at cosmological distances from the gravitational source.
When $\gamma _{W}$ is null (classical Schwarzschild solution), $k_{W}$ is
proportional to the cosmological scalar curvature of a de Sitter background $%
\cite{Mannheim 1989 Exact Solution}$.

\subsection{Mannheim-Kazanas parametrization}

\qquad When fitting the experimental galactic rotation curves with the
gravitational Weyl potential (Newtonian plus galactic term only), without
the assumption of any dark matter, Mannheim and Kazanas $\cite{Mannheim
1994 microlensing},\,\cite{Mannheim 1995 Age of Universe}$, 
$\cite{Mannheim 1989 Exact Solution}$ provided the following constraints on
the parameters $\beta _{W}$ and $\gamma _{W}$ of the theory: 
\begin{eqnarray}
\gamma _{W} &\sim &+10^{-26}\ \text{m}^{-1}\,,  \nonumber \\
\beta _{W}\text{ }(M_{Galaxy}) &\sim &+10^{+14}\ \text{m}^{+1}\,\text{.}
\label{parametrization_Mannheim_Kazanas}
\end{eqnarray}
On radial distances ($r$) smaller than the parsec, the ($\beta _{W}\ \gamma
_{W}$)- and $k_{W}$-contributions can be neglected in the effective
potential (\ref{effective_potential_WEYL}).

The explanation of the flattening of rotation curves on galactic distance
scales given by Mannheim and Kazanas requires $\gamma _{W}$ to be positive.
The order of magnitude of $\gamma _{W}$, close to the inverse of the Hubble
length, $c/H_{0}$, was noted as an interesting coincidence. However, we have
to keep in mind that the Mannheim-Kazanas parametrization (\ref
{parametrization_Mannheim_Kazanas}) is based on the specific choice\footnote{%
In a private communication $\cite{Mannheim 2002 conformal factor}$,
Mannheim advocated that ``all mass is dynamical'' in his theory. In other
words, both the test particle and the gravitational source would get their
mass from the same Higgs field, such that all massive particles possess the 
\emph{same} conformal factor, $\chi ^{2}(r),$ which makes it unobservable. 
\newline
However, it is known that there should exist at least two different symmetry
breaking mechanisms generating the ordinary matter mass. For example, the
electron gets its mass from the Higgs mechanism of the Standard Electroweak
Model, whereas the mass of the proton is mainly due to the Quantum
Chromodynamics. Hence, we claim that it is not allowed to attribute a common
conformal factor $\chi ^{2}(r)$ to all types of particles.} $\chi
^{2}(r)\equiv 1$, while any gravitational test or measurement requiring a
distance/time scale depends on the conformal factor.

\subsection{The weak field versus the strong field limit}

\qquad If we neglect the $k_{W}$-term, to which photons are blind as we
shall see, the radial distance scale at which the Weyl gravitational
potential in (\ref{effective_potential_WEYL}) can be considered to be a weak
field ($2\ V_{W}(r)/c^{2}<<1$) is given by 
\begin{equation}
\fl r_{weak\ field} 
\begin{array}[t]{cl}
\ll & 
\begin{array}[t]{l}
\frac{\pm 1+3\beta _{W}\ \gamma _{W}\pm \sqrt{1\pm \left\{ 
\begin{array}{c}
+14 \\ 
-2
\end{array}
\right\} \beta _{W}\ \gamma _{W}-3\left( \beta _{W}\ \gamma _{W}\right) ^{2}}%
}{2\gamma _{W}}
\end{array}
\medskip \\ 
\stackrel{\text{{\tiny (}}{\tiny \beta }_{{\tiny W}}{\tiny \gamma }_{{\tiny W%
}}\text{{\tiny )-term neglected}}}{\sim } & 
\begin{array}[t]{l}
\pm \frac{1}{\gamma _{W}}
\end{array}
\medskip \\ 
\stackrel{
\begin{array}{c}
\text{{\tiny if} }{\tiny \chi }^{{\tiny 2}}{\tiny (r)=1} \\ 
\text{{\tiny and}}{\tiny \,}\text{{\tiny (\ref
{parametrization_Mannheim_Kazanas})}}
\end{array}
}{\sim } & 
\begin{array}[t]{ll}
+10^{26}\ \text{m} & \,,
\end{array}
\end{array}
\label{weak_field_radius_Weyl}
\end{equation}

where $\beta _{W}\simeq G_{N}M/c^{2}$ with gravitational mass $M$; with the 
upper sign and number for $\gamma _{W}>0$ and the lower sign and number for $%
\gamma _{W}<0.$ When neglecting the ($\beta _{W}\ \gamma _{W}$)-term, the
weak field limit is independent of the gravitational mass creating the
field.\medskip

On the other hand, if distances are sufficiently large with respect to the
order of magnitude of $1/\left| \gamma _{W}\right| $\ and of the deflector
mass $M$, that is to say, 
\begin{equation}
\fl r_{strong\ field} 
\begin{array}[t]{cl}
\ggg & \sqrt{\frac{2\beta _{W}}{\left| \gamma _{W}\right| }}\medskip \\ 
\ggg & \sqrt{3} \cdot 10^{\frac{x+3}{2}}\frac{1}{\sqrt{\left| \gamma _{W}\right| }%
}\ \text{m for }M=10^{x}M_{Sun}\text{, }\gamma _{W}\text{ in }\left[ \text{m}%
^{-1}\right] \medskip \\ 
\stackrel{
\begin{array}{c}
\text{{\tiny if} }{\tiny \chi }^{{\tiny 2}}{\tiny (r)=1} \\ 
\text{{\tiny and}}{\tiny \,}\,\text{{\tiny (\ref
{parametrization_Mannheim_Kazanas})}}
\end{array}
}{\ggg } & \left\{ 
\begin{array}{l}
\left. 
\begin{array}{l}
2 \cdot 10^{20}\ \text{m for }M=10^{11}M_{Sun}
\end{array}
\right\} \text{\ a galaxy} \\ 
\left. 
\begin{array}{l}
2 \cdot 10^{21}\ \text{m for }M=10^{13}M_{Sun} \\ 
6 \cdot 10^{21}\ \text{m for }M=10^{14}M_{Sun} \\ 
2 \cdot 10^{22}\ \text{m for }M=10^{15}M_{Sun}
\end{array}
\right\} \text{ a cluster}\,,
\end{array}
\right.
\end{array}
\label{strong_field_radius_Weyl}
\end{equation}

the gravitational Weyl potential (\ref{effective_potential_WEYL}) can be
approximated to its linear and quadratic contributions only: 
\begin{equation}
V_{W}(r)_{\beta _{W}\equiv 0}=+\frac{\gamma _{W}}{2}\ r\ c^{2}-\frac{k_{W}}{2%
}\ r^{2}\ c^{2}\,.  \label{gravitational_potential_beta_null_Weyl}
\end{equation}
We first note that the strong field limit does depend on the mass source of
the gravitational field. Also, we note that, when $\beta _{W}\equiv 0$, the
radial variable $r$ must belong to certain allowed ranges to insure a proper
definition of the line element in Schwarzschild coordinates, $%
ds^{2}=A^{2}(r)\ c^{2}dt^{2}-B^{2}(r)\ dr^{2}-r^{2}\left[ d\theta ^{2}+\sin
^{2}\theta \ d\varphi ^{2}\right] $. In other words, the diagonal metric
terms in (\ref{metric_Schwarzschild_Weyl}) must be strictly positive. The
conditions are listed in discussion (\ref{condition_metric_Schwarzchild_Weyl}%
), section \ref{tables}. The relevance of the radial distances $r_{\min }$
and $r_{null}$ introduced there will become evident in the next
section.\medskip

Of course, one would need to analyze further the intermediate regime between
the weak field and strong field limits.

\section{Light deflection in Weyl theory}

\subsection{The geodesic equation}

\qquad \label{light_versus_massive}In the Weyl theory, the shape of the
effective geodesic potential for light crucially changes according to the
sign chosen for the parameter $\gamma _{W}$, and to whether the contribution
of some parameters ($\gamma _{W}$, $\beta _{W}$, $k_{W}$) is considered as
negligible or not. There might exist no black-hole solution. Moreover, the
effective geodesic potential for photons versus that for massive particles
presents different features that might enlighten us on the role of the
effective sign of the parameter $\gamma _{W}$, measured through light
deflection or galactic rotation curves. We can check those statements by
comparing the Weyl effective geodesic potential for light ($\digamma \equiv
0 $) with that for matter ($\digamma >0$) in Schwarzschild coordinates using
the metric given in equation (\ref{metric_Schwarzschild_Weyl}) with
definition (\ref{effective_potential_WEYL}). The geodesic equations can be
expressed as\newline
\begin{eqnarray}
\stackunder{\text{{\tiny ``Kinetic Energy''}}}{\underbrace{\left( \frac{dr}{%
d\lambda }\right) ^{2}}}+\stackunder{V_{\text{geodesic}}=\text{{\tiny %
``Effective potential''}}}{\underbrace{\left\{ \frac{1}{r^{2}}+\digamma 
\frac{\chi ^{2}(r)}{J^{2}}\right\} \left\{ 1+2\ \frac{V_{W}(r)}{c^{2}}%
\right\} }} &=&\stackunder{\text{{\tiny ``Total Energy''}}}{\underbrace{%
\frac{E^{2}}{J^{2}}}}  \label{effective_light/matter_potential_Weyl} \\
&&\smallskip  \nonumber
\end{eqnarray}
where $\frac{dr}{d\lambda }\equiv \frac{1}{r^{2}}\frac{dr}{d\varphi }$,
while $E$ and $J$ are integration constants for the total energy and the
total angular momentum of the particle respectively.\newline
First note that, as stated earlier, photon geodesics, unlike massive ones,
are independent of the conformal factor $\chi ^{2}(r)$, to be specified by
the symmetry breaking mechanism. This is why light deflection is a powerful
test for Weyl theory.\newline
To illustrate further the different behavior of massive versus massless
geodesics, let us fix arbitrarily $\chi ^{2}(r)$ to $1$ and consider the
effective gravitational force. The derivative of the effective geodesic
potential is given by $\cite{Edery 1998 weyl causal structure}$ 
\begin{equation}
-F_{\text{geodesic}}(r)\varpropto \frac{dV_{\text{geodesic}}}{dr}=\left. 
\begin{array}[t]{l}
-\frac{2}{r^{3}}\smallskip \\ 
+\beta _{W}\left( 2-3\beta _{W}\gamma _{W}\right) \left\{ \frac{3}{r^{4}}+%
\frac{\digamma }{J^{2}r^{2}}\right\} \smallskip \\ 
+3\beta _{W}\gamma _{W}\left\{ \frac{2}{r^{3}}\right\} \smallskip \\ 
+\gamma _{W}\left\{ -\frac{1}{r^{2}}+\frac{\digamma }{J^{2}}\right\}
\smallskip \\ 
+k_{W}\left\{ 0-2\frac{\digamma r}{J^{2}}\right\} \,\text{.}
\end{array}
\right.  \label{effective_light/matter_potential_derivative_Weyl}
\end{equation}
The interpretation of the first three terms contributing to the ``geodesic
force'' is unambiguous: the first term is asymptotically convergent with
respect to the variable $r$ and negative (repulsive) for any type of
particle. Whereas the factor corresponding to the Newtonian term
(multiplying $\beta _{W}\left( 2-3\beta _{W}\gamma _{W}\right) $) is always
positive (attractive), regardless of the type of particle. The same
conclusion holds for the factor in front of the ($\beta _{W}\gamma _{W}$%
)-term, which is always positive (thus attractive) if $\gamma _{W}$ is
positive, or alternatively negative (thus repulsive) for a negative $\gamma
_{W}$. Both the Newtonian term and this last term are asymptotically
convergent.\newline
On the contrary, the fifth term clearly distinguishes between massive and
nonmassive particles. The $k_{W}$-term has a null contribution for photons,
unlike for massive particles; and for nonrelativistic particles, it is
repulsive (negative) for a positive $k_{W}$ or conversely for a negative $%
k_{W}$. Moreover, the corresponding term in the potential, $V_{\text{geodesic%
}}(r)$, is asymptotically divergent for massive particles at large $r$.
However, in any case, we can neglect the $k_{W}$-contribution (set $k_{W}=0$%
) on non-cosmological distances.\newline
Let us therefore focus on the original feature of the Weyl gravity encoded
in the key parameter $\gamma _{W}$, namely the fourth term in (\ref
{effective_light/matter_potential_derivative_Weyl}). This term cannot be neglected at
intermediate distances. We realize that the sign of the factor corresponding to 
the parameter $\gamma _{W}$ depends on the type of particle: it is always
negative for photons ($\digamma =0$) and for sufficiently relativistic
particles, \emph{i.e.} for particles verifying 
\[
\left. 
\begin{array}{c}
\left\{ -\frac{1}{r^{2}}+\frac{\digamma }{J^{2}}\right\} <0\medskip \\ 
\begin{array}{lll}
\hspace{3.75cm} & \Updownarrow & \text{{\tiny if }}{\tiny dr/d\lambda \sim
0\ }\text{{\tiny in (\ref{effective_light/matter_potential_Weyl})}}
\end{array}
\medskip \\ 
J^{2}/E^{2}\gtrsim r_{0}^{2}/2\left\{ 1-2\ \frac{V_{W}(r)}{c^{2}}\right\}
\quad \text{for }V_{W}(r)/c^{2}\ \text{small.}
\end{array}
\right. 
\]
In contrast, it is positive for massive, not too relativistic particles. For
photons, moreover, the contribution of the $\gamma _{W}$-term in the
potential $V_{\text{geodesic}}(r)$ is convergent on large radial distances;
whereas it diverges linearly for massive, not too relativistic particles.
Again, for photons, this term is attractive for $\gamma _{W}<0$ or repulsive
for $\gamma _{W}>0$; while it is just the opposite for nonrelativistic
particles.

\subsection{Critical radii from the effective geodesic potential for photons}

\qquad We now analyze thoroughly the case of null geodesics \textbf{\ }($%
\digamma \equiv 0$ and we set $k_{W}$ to zero\textbf{)}, that is, the
trajectory of photons and ultra relativistic particles. We find the
following characteristics for the effective geodesic light potential: a
radius at which the effective geodesic light potential is minimal, one at
which it shows a local maximum, one being an inflection point, and finally,
a radius at which the effective geodesic light potential cancels. We can
approximate those critical radii assuming that $\beta _{W}\gamma _{W}$ is
small in comparison with $\beta _{W}$ and $\gamma _{W}$ alone. Their
expressions are respectively,\newline
\begin{equation}
\left. 
\begin{array}{l}
r_{\min }=\frac{3\beta _{W}\gamma _{W}-2}{\gamma _{W}}\sim -\frac{2}{\gamma
_{W}}\,,\medskip \\ 
r_{\max }=3\beta _{W}\,,\medskip \\ 
r_{\text{inflection}}=\frac{-3+9\beta _{W}\gamma _{W}-\sqrt{9-54\beta
_{W}\gamma _{W}+81\left( \beta _{W}\gamma _{W}\right) ^{2}+48\gamma
_{W}-72\beta _{W}\gamma _{W}^{2}}}{2\gamma _{W}}\sim -\frac{3}{\gamma _{W}}%
\,,\medskip \\ 
r_{null}=\frac{-1+3\beta _{W}\gamma _{W}-\sqrt{1+2\beta _{W}\gamma
_{W}-3\left( \beta _{W}\gamma _{W}\right) ^{2}}}{2\gamma _{W}}\sim -\frac{1}{%
\gamma _{W}}\,.\medskip
\end{array}
\right.  \label{important_radius_Weyl}
\end{equation}

Analyzing the shape of the effective geodesic potential curve for light in
some limiting cases will provide some insight on the key radial distances
introduced here.\medskip

\textbf{Case A. }$\mathbf{\gamma }_{\mathbf{W}}\mathbf{=0}$\smallskip

When $\gamma _{W}$ is null, we essentially recover General Relativity: the
effective geodesic light potential is that of a Schwarzschild black hole
with a generalized Schwarzschild radius $\overline{r}$ (Figure \ref
{v_geodesic_gamma_null}). There thus exists a critical value of the closest
approach distance, $r_{\max }$, under which no deflection can take place
because photons are captured. Hence, the condition for a photon on a
trajectory $r(t)$ to be deflected with a closest approach distance $r_{0}$
to the gravitational mass is: 
\begin{equation}
r\geq r_{0}>r_{\max }>\overline{r}\equiv 2\beta _{W}\ .
\label{r0_critique_all_gamma_Weyl}
\end{equation}
\medskip

\textbf{Case B. }$\mathbf{\beta }_{\mathbf{W}}\mathbf{=0}$\smallskip

When $\beta _{W}$ is assumed to be zero, the shape of the potential for
light depends crucially on the sign of $\gamma _{W}$.

\qquad \quad \textbf{B1. }$\mathbf{\gamma }_{\mathbf{W}}\mathbf{>0}$

For positive values of $\gamma _{W}$, the effective geodesic potential for
light is a wall (Figure \ref{v_geodesic_gamma_positive}), and thus there
exists no critical value of the closest approach distance. Indeed, the
characteristic points of the curve (minimum, inflection point) are at
negative radii and thus are non physical. Light deflection can occur at any
radial distance and is characterized by 
\begin{equation}
\text{for }\gamma _{W}>0:\quad r\geq r_{0}\,.
\label{r0_critique_gamma_positive_Weyl}
\end{equation}

\qquad \quad \textbf{B2. }$\mathbf{\gamma }_{\mathbf{W}}\mathbf{<0}$

In case of a negative $\gamma _{W}$, we have, again, a potential wall with
no critical value of the closest approach distance, but it admits a minimum
and negative energies (Figure \ref{v_geodesic_gamma_negative}). This $%
r_{\min }$ constitutes the radius of a stable circular orbit for photons,
and together with other orbits of negative effective total energy ($\left. 
\frac{E^{2}}{J^{2}}\right| _{photon}<0$) it belongs to the class of bound
states. There thus exists a maximum closest approach distance, $r_{null}$,
allowed for asymptotically free trajectories. Consequently, the condition
for light deflection to take place is: 
\begin{equation}
\text{for }\gamma _{W}<0:\quad r\geq r_{0}\quad \text{and }r_{0}\leq
r_{null}\,.  \label{r0_critique_gamma_negative_Weyl}
\end{equation}
The role of $r_{null}$ will be clarified in the next paragraph, particularly
with expression (\ref{orbits_Weyl}).\medskip

\textbf{Case C. Nonzero values of }$\mathbf{\gamma }_{\mathbf{W}}$\textbf{\
and }$\mathbf{\beta }_{\mathbf{W}}$\smallskip

On short distance scales ($r<<1/\left| \gamma _{W}\right| $), the potential
is Schwarzschild-like and analogous to case A discussed above.\newline
On intermediate distances ($r\sim 1/\left| \gamma _{W}\right| $), if $\gamma
_{W}<0$, the effective geodesic potential for light presents a minimum as
described in case B2. Otherwise, for $\gamma _{W}>0$, the comments given in
case B1 above apply.

\subsection{Conditions for light deflection}

From the geodesic equations for photons, one derives the photon path,
parametrized as the angle $\varphi $ as a function of the radial distance $r$
to the gravitational source. For a generic metric in Schwarzschild
coordinates ($r$, $\theta $, $\varphi $):

\begin{equation}
\varphi (r)=\int_{0}^{\sin \Phi =r_{0}/r}\frac{\cos \Phi }{\sqrt{%
A^{2}(r_{0})-\sin ^{2}\Phi \ A^{2}(r)}}\ d\Phi \,.
\label{trajectory_exact_Schwarzschild}
\end{equation}

The asymptotic light deflection angle, $\hat{\alpha}$, is usually defined as
the angle between the inner and outer asymptotes to the photon trajectory,
with $r_{0}$ the closest approach distance to the gravitational body (Figure 
\ref{asymptotic_deflexion_lumiere_def}): 
\begin{equation}
\hat{\alpha}_{\text{exact}}(r_{0})=2\left| \varphi (r_{0})\right| -\pi \ .
\label{asymptotic_deflection_angle}
\end{equation}

\subsubsection{The weak field regime}

\paragraph{The light deflection angle}

\quad \newline
Let us apply the general expression for the photon path (\ref
{trajectory_exact_Schwarzschild}) and the asymptotic deflection angle (\ref
{asymptotic_deflection_angle}) to the Weyl metric (\ref
{metric_Schwarzschild_Weyl}) in the weak field regime (\ref
{weak_field_radius_Weyl}). We obtain: 
\begin{eqnarray}
\hat{\alpha}_{\text{weak field}}(r_{0}) &\simeq &+\frac{2\beta _{W}\ \left(
2-3\beta _{W}\ \gamma _{W}\ \right) }{r_{0}}+\frac{3}{2}\ \beta _{W}\ \gamma
_{W}\ \ \pi -\gamma _{W}\ \ r_{0}\,,  \label{angle_deflex_weak_WEYL} \\
&&\text{where}\ r_{0}\text{ verifies (\ref{weak_field_radius_Weyl}) with} 
\begin{array}[t]{l}
\text{(\ref{r0_critique_all_gamma_Weyl}), if }\gamma _{W}\geq 0\,, \\ 
\text{(\ref{r0_critique_all_gamma_Weyl}) and (\ref
{r0_critique_gamma_negative_Weyl}), if }\gamma _{W}<0\,,
\end{array}
\nonumber
\end{eqnarray}
\newline
We see that the ($\beta _{W}\ \gamma _{W}$)-term contributes to a small
constant deflection, independent of the radial distance, while the linear
term tends to decrease or increase the usual Newtonian contribution (first
term in (\ref{angle_deflex_weak_WEYL})), according to the sign of the
parameter $\gamma _{W}$.

\paragraph{Critical radius from the weak field deflection angle}

\quad \newline
If $\gamma _{W}$ is negative, the asymptotic deflection angle in (\ref
{angle_deflex_weak_WEYL}) is always convergent (positive); whereas if $%
\gamma _{W}$ is positive, the deflection angle cancels at $r_{0\ 0}$, which
is a function of the deflector mass, and becomes divergent (negative) beyond
that distance: 
\begin{equation}
\fl r_{0\ 0}\left. 
\begin{array}[t]{cl}
\equiv & \frac{3\pi \ \beta _{W}\ \gamma _{W}+\sqrt{9\ \left( \beta _{W}\
\gamma _{W}\right) ^{2}\pi ^{2}+32\ \beta _{W}\ \gamma _{W}\left( 2-3\ \beta
_{W}\ \gamma _{W}\right) }}{4\gamma _{W}}\smallskip \\ 
\stackrel{\text{{\tiny (}}{\tiny \beta }_{{\tiny W}}{\tiny \gamma }_{{\tiny W%
}}\text{{\tiny )-term neglected}}}{\sim } & 2\sqrt{\frac{\beta _{W}}{\gamma
_{W}}}\smallskip \\ 
\sim & \sqrt{6} \cdot 10^{\frac{x+3}{2}}\frac{1}{\sqrt{\left| \gamma _{W}\right| }%
}\ \text{m for }M=10^{x}M_{Sun}\text{, }\gamma _{W}\text{ in }\left[ \text{m}%
^{-1}\right] \medskip \\ 
\stackrel{
\begin{array}{c}
\text{{\tiny if} }{\tiny \chi }^{{\tiny 2}}{\tiny (r)=1} \\ 
\text{{\tiny and}}{\tiny \,}\,\text{{\tiny (\ref
{parametrization_Mannheim_Kazanas})}}
\end{array}
}{\sim } & \left\{ 
\begin{array}{l}
\left. 
\begin{array}{l}
8 \cdot 10^{14}\ \text{m for }M=M_{Sun}\,\text{,}
\end{array}
\right. \smallskip \\ 
\left. 
\begin{array}{l}
2 \cdot 10^{20}\ \text{m for }M=10^{11}M_{Sun}\,\text{,}
\end{array}
\right\} \text{ a galaxy}\smallskip \\ 
\left. 
\begin{array}{l}
2 \cdot 10^{21}\ \text{m for }M=10^{13}M_{Sun}\,\text{,}\smallskip \\ 
8 \cdot 10^{21}\ \text{m for }M=10^{14}M_{Sun}\,\text{,}\smallskip \\ 
2 \cdot 10^{22}\ \text{m for }M=10^{15}M_{Sun}\,\text{,}
\end{array}
\right\} \text{ a cluster}\,.
\end{array}
\right.
\end{array}
\right. \smallskip  \label{r0_0_Weyl}
\end{equation}

\subsubsection{The strong field regime}

\paragraph{Open versus closed orbits in the strong field regime}

\quad \newline
If we apply the general expression for the photon path (\ref
{trajectory_exact_Schwarzschild}) to the Weyl metric (\ref
{metric_Schwarzschild_Weyl}) in the strong field regime (\ref
{strong_field_radius_Weyl}, \ref{gravitational_potential_beta_null_Weyl}),
we see that we need additional conditions on the distances $r$ and $r_{0}$,
so that the term in the square-root of the integrand of equation 
(\ref{trajectory_exact_Schwarzschild}), which we call $\Delta (r)$, 
be positive \footnote{%
It should be further restricted to a strict inequality for the integration
variable $r$ in expression (\ref{trajectory_exact_Schwarzschild}).} 
when $\beta _{W}= 0$. 
Those conditions, together with the definition of $\Delta _{\beta _{W}\equiv 0}(r)$, 
are listed in discussion (\ref{condition_delta_positive_Weyl}),
section \ref{tables}.\medskip

Bearing discussions (\ref{condition_metric_Schwarzchild_Weyl}) and (\ref
{condition_delta_positive_Weyl}) from section \ref{tables} in mind, the
potential in the strong field regime (\ref
{gravitational_potential_beta_null_Weyl}) permits an exact integration of
the photon trajectory (\ref{trajectory_exact_Schwarzschild}) given by 
\begin{eqnarray*}
\varphi (r)_{\beta _{W}\equiv 0} &=&\left[ \pm \arctan \left( \frac{r_{0}\
\left( \gamma _{W}\ r+2\right) }{2\sqrt{\Delta (r)}}\right) \right]
_{r_{initial}}^{r}\medskip \\
&& 
\begin{array}{l}
\medskip \\ 
\text{with }\left\{ 
\begin{array}{ll}
r_{initial}=+\infty & \text{for an unbound orbit\smallskip } \\ 
r_{initial}=r_{0} & \text{for a bound orbit}
\end{array}
\right.
\end{array}
\end{eqnarray*}
\begin{eqnarray}
&\Rightarrow &\pm \ \varphi (r)_{\beta _{W}\equiv 0}\ \pm \ \varphi
_{initial}=\arcsin \left( \frac{2\ r_{0}/r+\gamma _{W}\ r_{0}}{2+\gamma
_{W}\ r_{0}}\right) \text{\smallskip }  \nonumber \\
&\Leftrightarrow &r_{\beta _{W}\equiv 0}=\frac{-2/\gamma _{W}}{1-\frac{%
2+\gamma _{W}\ r_{0}}{\gamma _{W}\ r_{0}}\cdot \sin \left( \pm \varphi \pm
\varphi _{initial}\right) }  \label{orbit_radius_weyl} \\
&& 
\begin{array}{l}
\medskip \\ 
\text{where } \\ 
\begin{array}{l}
\varphi _{initial}\equiv \left\{ 
\begin{array}{l}
+\arcsin \left( \frac{\gamma _{W}\ r_{0}}{2+\gamma _{W}\ r_{0}}\right) \quad
\left. 
\begin{array}[t]{l}
\forall \text{ }r_{0}\text{ if }\gamma _{W}>0\,\text{;\smallskip } \\ 
\text{or for }r_{0}<r_{null}\text{ if }\gamma _{W}>0\text{\smallskip }
\end{array}
\right. \\ 
\mp \frac{\pi }{2}\quad \text{for }r_{0}=r_{null}\text{ or }r_{0}>r_{\min }%
\text{ if }\gamma _{W}<0\text{\smallskip } \\ 
\pm \frac{\pi }{2}\quad \text{for }r_{null}<r_{0}<r_{\min }\text{ if }\gamma
_{W}<0
\end{array}
\right. \medskip \\ 
e\equiv \left| \frac{2+\gamma _{W}\ r_{0}}{\gamma _{W}\ r_{0}}\right| \,%
\text{the eccentricity}.
\end{array}
\end{array}
\nonumber
\end{eqnarray}
\smallskip \newline
The types of allowed orbits can thus be classified $\cite{Edery 1998 weyl
causal structure}$ according to (\ref{orbits_Weyl}), with respect to $%
r_{null}$, $r_{\min }$ and $r_{*}$ given in (\ref{important_radius_Weyl})
and (\ref{condition_delta_positive_Weyl}). Figures \ref
{bis_photon_trajectories_weyl} and \ref{photon_trajectories_weyl} illustrate
the different photon orbits.\bigskip 
\begin{equation}
\fl
\fbox{$\left. 
\begin{array}{c}
types\ of\ orbits\ in\ Schwarzschild\ coordinates:\vspace{1cm} \\ 
\left. 
\begin{array}{l}
\text{for}\ r\text{ verifying (\ref{condition_metric_Schwarzchild_Weyl}) and
(\ref{condition_delta_positive_Weyl}):\medskip } \\ 
\left. 
\begin{array}{l}
\quad \text{if }\gamma _{W}>0\,\text{:}\quad \forall \ r_{0}\text{,\quad
hyperbolic} \\ 
\quad \text{if }\gamma _{W}<0\,\text{:}\quad 
\begin{array}[t]{l}
r_{0}<r_{null}\text{,\quad hyperbolic (}e>1\text{)} \\ 
r_{0}=r_{null}\text{,\quad parabolic\hspace{0.3cm}(}e=1\text{)} \\ 
r_{0}>r_{null}\text{,\quad elliptic\hspace{0.7cm}(}e<1\text{)} \\ 
\hspace{0.9cm}\text{with perihelion/aphelion given by }r_{0}\text{ or }r_{*}
\\ 
\hspace{0.9cm}\text{particular case if }r_{0}=r_{\min }\text{,\quad circular
(}e=0\text{)\thinspace .}
\end{array}
\end{array}
\right.
\end{array}
\right.
\end{array}
\right. $}  \label{orbits_Weyl}
\end{equation}
\smallskip \smallskip

We see that, in fact, conditions (\ref{condition_metric_Schwarzchild_Weyl})
only reflect the different orbits allowed for photons.

Note that, strictly speaking, expression (\ref{orbit_radius_weyl}) cannot be
used for the circular orbit. Instead, one needs to come back to the geodesic
equation for photons ($\digamma =0$) (\ref
{effective_light/matter_potential_Weyl}) and to its derivative with respect
to $r$. Then, constant radius orbits are found when imposing $\partial
r/\partial \lambda =\partial ^{2}r/\partial ^{2}\lambda =0$. The circular
orbit radii obtained correspond to $r_{\min }$ and $r_{\max }$ given in (\ref
{important_radius_Weyl}), of which $r_{\max }$ is unstable.

\paragraph{The light deflection angle}

\quad \newline
Consequently, in Weyl theory, light deflection in the strong field regime is
described by the following exact equation 
\begin{eqnarray}
\hat{\alpha}_{\beta _{W}\equiv 0}(r_{0}) &=&-2\arcsin \left( \frac{\gamma
_{W}\ r_{0}}{2+\gamma _{W}\ r_{0}}\right) \,,\medskip
\label{angle_deflex_exact_large_distance_WEYL} \\
&&\text{where}\ r_{0}\text{ verifies (\ref{strong_field_radius_Weyl}) and } 
\begin{array}[t]{l}
\text{(\ref{r0_critique_gamma_positive_Weyl}), if }\gamma _{W}>0 \\ 
\text{(\ref{r0_critique_gamma_negative_Weyl}), if }\gamma _{W}<0\ .
\end{array}
\nonumber
\end{eqnarray}
It corresponds to asymptotically free orbits (hyperbolic or parabolic)
described in discussion (\ref{orbits_Weyl}).

\section{Amazing features of the strong field limit for a negative value of $%
\gamma _{W}$}

\qquad Let us now get a visual overview on how the various critical radii
introduced here structure the space-time with respect to light paths, and
hence structure information transfer.

\subsection{An accumulation point}

\qquad To start with, we realize from the graph of photon trajectories
(Figure \ref{photon_trajectories_weyl}) that the semi-lattice rectum,
located at $r=-2/\gamma _{W}$ for \emph{any} orbit in the strong field
limit, is an accumulation point. Indeed, if we interpolate between the weak
field regime, in which the description of light deflection is qualitatively
analogous to the description in General Relativity, and the strong field
regime, we might possibly find the picture illustrated in Figure \ref
{accumulation_point_weyl}. Consequently, the surface of a sphere of radius $%
r=-2/\gamma _{W}$, centered on a given lens, is a very particular region of
space.

\subsection{Peculiar alignment configuration}

\qquad Another remark is the existence of a particular Observer-Lens-Source
alignment configuration in the strict strong field regime ($\beta _{W}=0$),
in which the observer and the source are located on the semi-lattice rectum
points. There would then be an infinite number of photon trajectories coming
from the source to the observer (see Figure \ref{particular_ols_weyl})!

\subsection{Observable regions of space}

\qquad It would also be interesting to investigate further which regions of
space are observable (connected by light geodesics) in the strong field
regime. For example, in the presence of one single lens (L), when we wish to
understand if an observer, in the strict strong field limit ($\beta _{W}=0$%
), will be able to see some given source, we realize that it depends on the
position of the source with respect to the concentric spheres of radius $%
r_{null}$ and $r_{\min }$ surrounding the lens. The equation of orbits
described in (\ref{orbit_radius_weyl}) and (\ref{orbits_Weyl}) are useful
for such a discussion. Their consequences are summarized in Figure \ref
{visibility_ols_weyl_12345}.

Indeed, if the light source (S$_{1}$) is inside the smallest sphere of
radius $r_{null}$, then, the closest approach distance of any photon
originating from the source is necessarily smaller than $r_{null}$. This
means that all these photons have unbound orbits (hyperbolas centered on L).%
\newline
Also, if the source (S$_{2}$) is on the sphere of radius $r_{null}$, the
closest approach distance of its photons can be either smaller or equal to $%
r_{null}$, which means that all the orbits are again unbound (hyperbolas or
a parabola). In those two cases (S$_{1}$ and S$_{2}$), the source can be
seen from any region of space (except just behind the lens if it is not
transparent). Whereas if the source (S$_{3}$) is located in between the two
spheres, there are in addition photons originating from this source that
have a closest approach distance to the lens larger than $r_{null}$.
Accordingly, those photons are captured on elliptical orbits. Hence, in this
case, there will be some regions of space that cannot be reached by any
photon of S$_{3}$.\newline
We have already evoked the particular case of a source (S$_{4}$) situated on
the sphere of radius $r_{\min }$, which allows all four types of orbits:
hyperbolic, parabolic, circular and elliptic.\newline
For a source outside the larger sphere (S$_{5}$) there will also be some
regions of space that cannot be reached by any photon originating from the
source.\newline
One could now complicate the game by considering many lenses and relax the
strong field regime!

\section{Conclusions}

\qquad We have shown how in Weyl theory different critical radial distances
structure the space-time surrounding a gravitational mass with respect to
photon paths. This structure and the physical relevance of the different
radii (relevant when positive!) strongly depend on the sign of the key
parameter of the theory, $\gamma _{W}$.

\emph{When }$\gamma _{W} $\emph{\ is negative}, it is important to notice the
coincidence between the weak field limiting radius (\ref
{trajectory_exact_Schwarzschild}) and the critical radius $r_{null}$ (\ref
{important_radius_Weyl}) that separates between open and closed orbits; in
other words, between the possibility and the impossibility of light
deflection. This means that in the weak field limit, that is when the
Newtonian contribution dominates, light deflection always takes place. On
the contrary, in the strong field regime, when the $\gamma _{W}$-term is
nonnegligible, light deflection is not always possible.\newline
If we consider the order of magnitude given for $\gamma _{W}$ by Mannheim
and Kazanas, resting on a particular scale factor, then $r_{null}\sim 
10^{26}$ m is a cosmological distance. On the other hand, if we consider
the more conservative estimate of $\gamma _{W}$ derived in article $\cite
{Pireaux 2003 constraints on Weyl parameter}$, then $r_{null}$ $\gtrsim
10^{18}$ m might be within galactic distances. It is important to note that
this later estimate does not imply any restriction on the conformal factor.

\emph{When }$\gamma _{W}$\emph{\ is positive}, light deflection always takes
place, whether the weak field regime be valid or not. However, a new feature
of Weyl gravity, by comparison to General Relativity, appears at closest
approach distances larger than the critical value r$_{0 \ 0}$ (\ref{r0_0_Weyl}%
): light deflection from being a convergent phenomenon becomes divergent. In
the strong field regime, light deflection is always divergent because the
order of magnitude of the strong field limiting radius (\ref
{strong_field_radius_Weyl}) is the same as that of the critical closest
approach distance (\ref{r0_0_Weyl}).
\pagebreak

\section{TABLES AND FIGURES}

\label{tables}\bigskip

\FRAME{fhFU}{3.9643in}{2.092in}{0pt}{\Qcb{Definition of the asymptotic light
deflection angle. I= Image; S= Source; O= Observer; L= Lens (gravitational
potential); $\widehat{\alpha }$= asymptotic light deflexion angle; $r_{0}$=
closest approach distance, written here in Schwarzschild coordinates ($r$, $%
\theta $, $\varphi $); $b$= impact parameter. The solid line shows the light
path while the dash line shows the inner and outer asymptotes to the
geodesics.}}{\Qlb{asymptotic_deflexion_lumiere_def}}{%
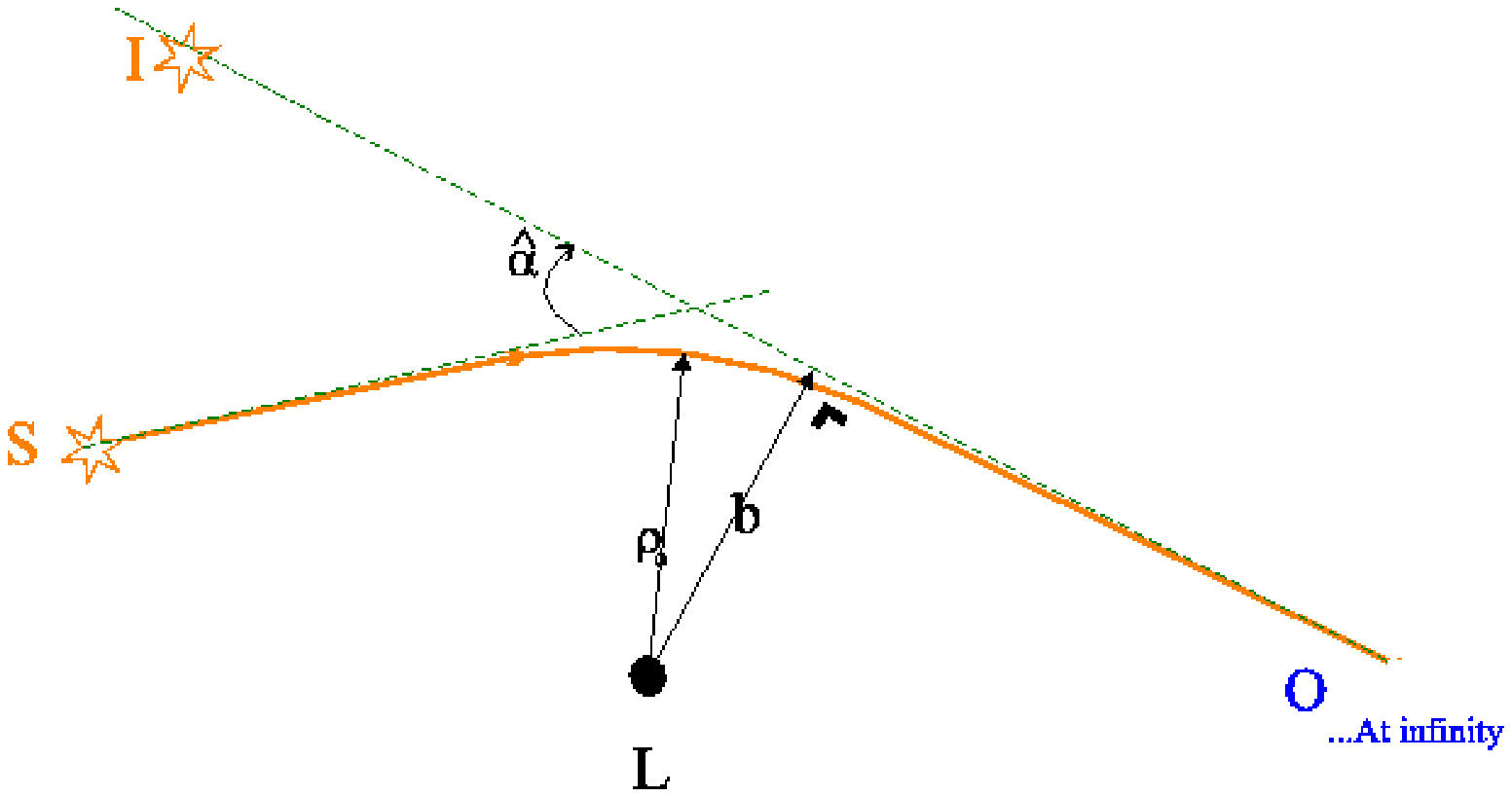}{\special{language "Scientific
Word";type "GRAPHIC";maintain-aspect-ratio TRUE;display "ICON";valid_file
"F";width 3.9643in;height 2.092in;depth 0pt;original-width
0pt;original-height 0pt;cropleft "0";croptop "1";cropright "1";cropbottom
"0";filename
'asymptotic_deflexion_lumiere_def.eps';file-properties "XNPEU";}%
}
\pagebreak

\begin{equation}
\fl
\fbox{$\left. 
\begin{array}{c}
proper\ definition\ of\ the\text{ }Weyl\text{ }line\text{ }element\text{ }in%
\text{ }Schwarzschild\text{ }coordinates\medskip \\ 
\Rightarrow A_{\beta _{W}\equiv 0}^{2}(r)=B_{\beta _{W}\equiv
0}^{-2}(r)=\left[ 1+\frac{2\ V_{W}}{c^{2}}\right] >0\quad with\text{ }(\ref
{gravitational_potential_beta_null_Weyl})\text{:}\vspace{1cm} \\ 
\left. 
\begin{array}{l}
\text{Let }K\equiv -k_{W}-\frac{\gamma _{W}^{2}}{4}\,\text{,\bigskip } \\ 
\;\text{if }K>0\Leftrightarrow k_{W}<-\frac{\gamma _{W}^{2}}{4}<0\,\text{%
:\quad }r\in \left[ 0,+\infty \right[ \medskip \\ 
\;\text{if }K=0\Leftrightarrow k_{W}=-\frac{\gamma _{W}^{2}}{4}\,\text{: }%
\left\{ 
\begin{array}{l}
\gamma _{W}>0\Rightarrow r_{\min }<0\,\text{:\quad }r\in \left[ 0,+\infty
\right[ \smallskip \\ 
\gamma _{W}<0\Rightarrow r_{\min }>0\,\text{:\quad }r\in \left[ 0,r_{\min
}\right[
\end{array}
\right. \medskip \\ 
\;\text{if }K<0\,\text{: }\left\{ 
\begin{array}{l}
k_{W}>0\Rightarrow r_{-}<0\text{ and }r_{+}>0\,\text{:\quad }r\in \left[
0,r_{+}\right[ \smallskip \\ 
k_{W}=0\,\text{: }\left\{ 
\begin{array}{l}
\gamma _{W}>0\Rightarrow r_{null}<0\,\text{:\quad }r\in \left[ 0,+\infty
\right[ \smallskip \\ 
\gamma _{W}<0\Rightarrow r_{null}>0\,\text{:\quad }r\in \left[
0,r_{null}\right[
\end{array}
\right. \smallskip \\ 
k_{W}<0\,\text{: }\left\{ 
\begin{array}{l}
\gamma _{W}>0\Rightarrow r_{-}<0\text{ and }r_{+}<0\,\text{:\quad }r\in
\left[ 0,+\infty \right[ \smallskip \\ 
\gamma _{W}<0\Rightarrow r_{-}>r_{+}>0\,\text{:\quad }r\in \left[
0,r_{+}\right[ \cup \left] r_{-},+\infty \right[
\end{array}
\right.
\end{array}
\right. \bigskip \\ 
\;\text{where }\left. 
\begin{array}[t]{l}
r_{\min }\text{ is the root of }A_{\beta _{W}\equiv 0}^{2}(r)\text{ when }%
K=0\smallskip \\ 
r_{\pm }\equiv \frac{\gamma _{W}\pm \sqrt{-4K}}{2\ k_{W}}\text{ are the
roots of }A_{\beta _{W}\equiv 0}^{2}(r)\text{ when }K<0\smallskip \\ 
r_{null}\text{ is the root of }A_{\beta _{W}\equiv 0}^{2}(r)\text{ when }%
k_{W}=0\smallskip \\ 
\text{... }\nexists \ \text{roots of }A_{\beta _{W}\equiv 0}^{2}(r)\text{
when }K>0\,.
\end{array}
\right.
\end{array}
\right.
\end{array}
\right. $}	\label{condition_metric_Schwarzchild_Weyl}
\end{equation}

\begin{equation}
\fl
\fbox{$\left. 
\begin{array}{c}
conditions\ for\ \Delta _{\beta _{W}\equiv 0}(r)\equiv \left( 1+\gamma _{W}\
r_{0}\right) r^{2}-\gamma _{W}\ r_{0}^{2}\ r-r_{0}^{2}\geq 0\ in\ (\ref
{trajectory_exact_Schwarzschild}):\vspace{1cm} \\ 
\left. 
\begin{array}{l}
\text{if }\gamma _{W}>0\,\text{:\quad }r\in \left[ r_{0},+\infty \right[
\medskip \\ 
\text{if }\gamma _{W}<0\,\text{: }\left\{ 
\begin{array}{l}
0<r_{0}<r_{null}\Rightarrow r_{*}<0\,\text{:\quad }r\in \left[ r_{0},+\infty
\right[ \smallskip \\ 
r_{0}=r_{null}\,\text{:\quad }r\in \left[ r_{0},+\infty \right[ \smallskip
\\ 
r_{null}<r_{0}<r_{\min }\Rightarrow 0<r_{0}<r_{*}\,\text{: \quad }r\in
\left[ r_{0},r_{*}\right] \smallskip \\ 
r_{0}=r_{\min }\text{ }\Rightarrow r_{*}=r_{0}\smallskip \\ 
r_{0}>r_{\min }\Rightarrow 0<r_{*}<r_{0}\,\text{: \quad }r\in \left[
r_{*},r_{0}\right]
\end{array}
\right. \bigskip \\ 
\left. 
\begin{array}[t]{l}
\text{where }\smallskip \\ 
\;\left. 
\begin{array}{l}
r_{*}\equiv \frac{-r_{0}}{1+\gamma _{W}\ r_{0}} \\ 
r_{0}
\end{array}
\right\} \text{ are the roots of }\Delta _{\beta _{W}\equiv 0}(r)\text{ when 
}r_{0}<r_{\min }\text{ and }r_{0}\neq r_{null}\smallskip \\ 
\;\left. r_{null}\text{ is the root of }\Delta _{\beta _{W}\equiv 0}(r)\text{
when }r_{0}=r_{null}\right. \smallskip \\ 
\;\left. \nexists \ \text{roots of }\Delta _{\beta _{W}\equiv 0}(r)\text{
when }r_{0}>r_{\min }\text{ and }\gamma _{W}<0\,.\right.
\end{array}
\right.
\end{array}
\right.
\end{array}
\right. $}	\label{condition_delta_positive_Weyl}
\end{equation}
\pagebreak

\FRAME{fhF}{5.3949cm}{4.971cm}{0cm}{\Qcb{Effective geodesic potential for
light when $\gamma _{W}$ is null.}}{\Qlb{v_geodesic_gamma_null}}{%
v_geodesic_gamma_null.eps}{\special{language "Scientific Word";type
"GRAPHIC";maintain-aspect-ratio TRUE;display "ICON";valid_file "F";width
5.3949cm;height 4.971cm;depth 0cm;original-width 0pt;original-height
0pt;cropleft "0";croptop "1";cropright "1";cropbottom "0";filename
'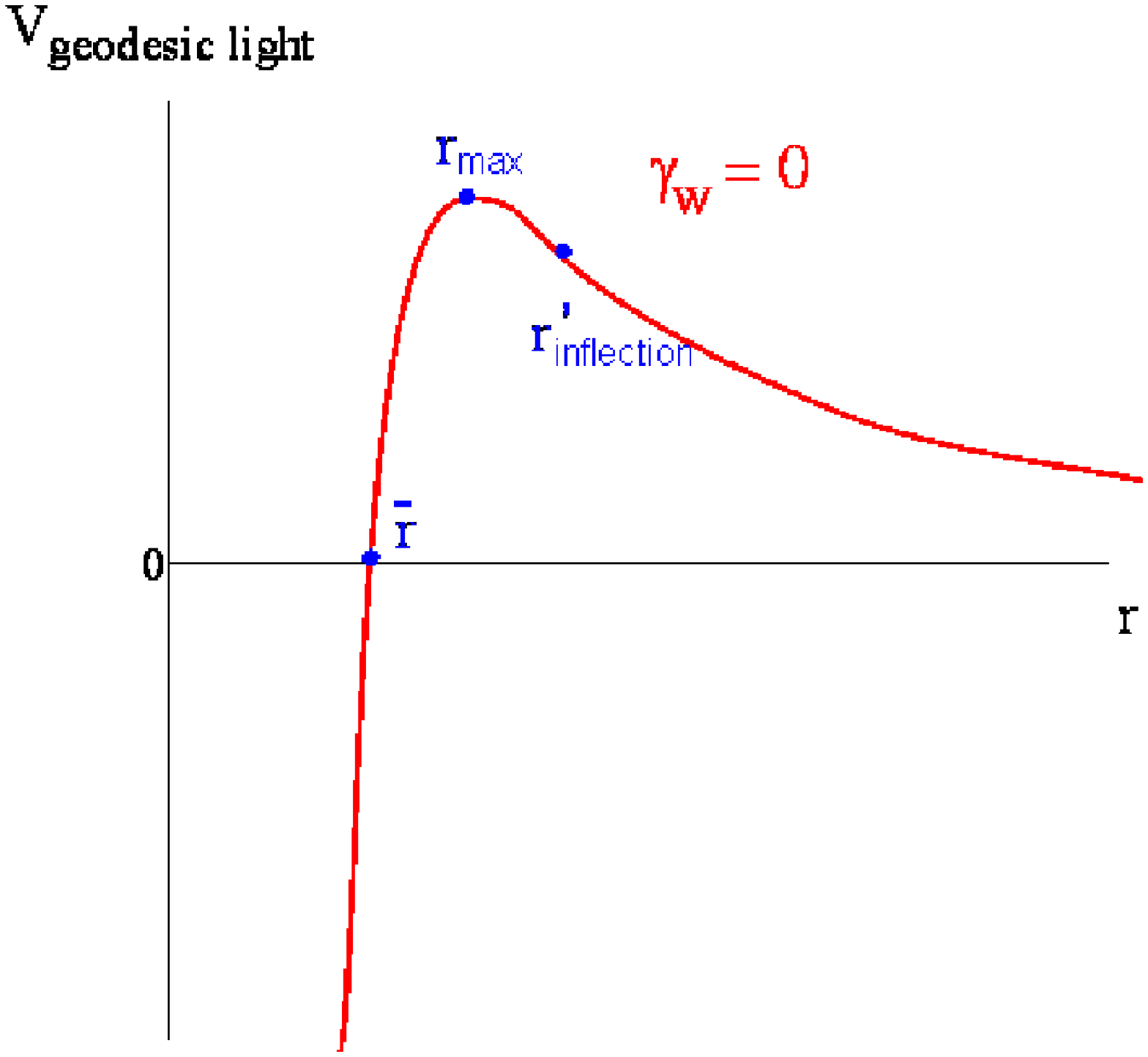';file-properties "XNPEU";}%
}

\FRAME{fhF}{2.1162in}{1.9363in}{0in}{\Qcb{Effective geodesic potential for
light when $\gamma _{W}$ is positive and $\beta _{W}$ is null.}}{\Qlb{%
v_geodesic_gamma_positive}}{v_geodesic_gamma_positive.eps}{\special{language
"Scientific Word";type "GRAPHIC";maintain-aspect-ratio TRUE;display
"ICON";valid_file "F";width 2.1162in;height 1.9363in;depth
0in;original-width 0pt;original-height 0pt;cropleft "0";croptop
"1";cropright "1";cropbottom "0";filename
'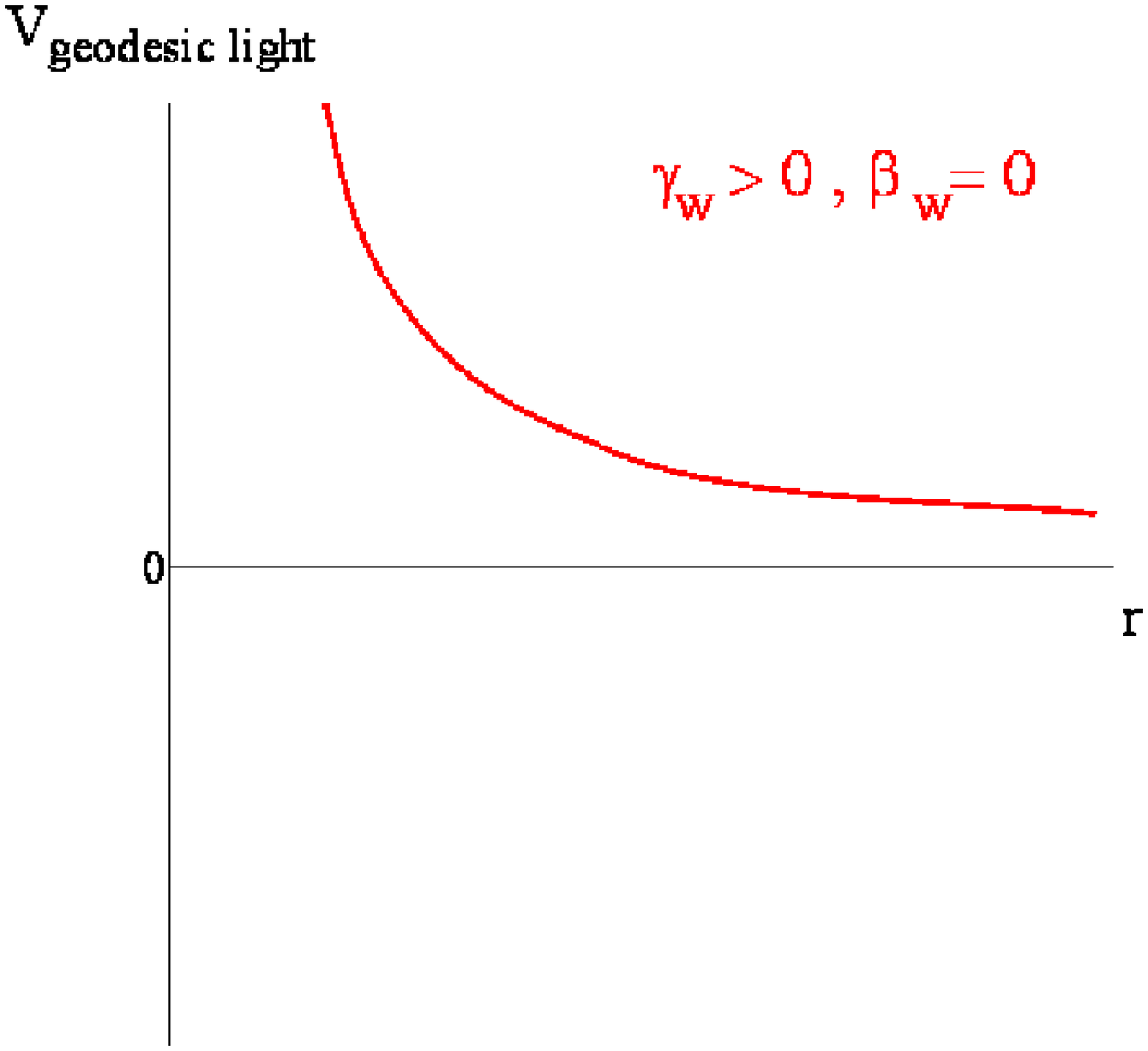';file-properties "XNPEU";}%
}

\FRAME{fhFU}{2.1162in}{1.9363in}{0pt}{\Qcb{Effective geodesic potential for
light when $\gamma _{W}$ is negative and $\beta _{W}$ is null.}}{\Qlb{%
v_geodesic_gamma_negative}}{v_geodesic_gamma_negative.eps}{\special{language
"Scientific Word";type "GRAPHIC";maintain-aspect-ratio TRUE;display
"ICON";valid_file "F";width 2.1162in;height 1.9363in;depth
0pt;original-width 0pt;original-height 0pt;cropleft "0";croptop
"1";cropright "1";cropbottom "0";filename
'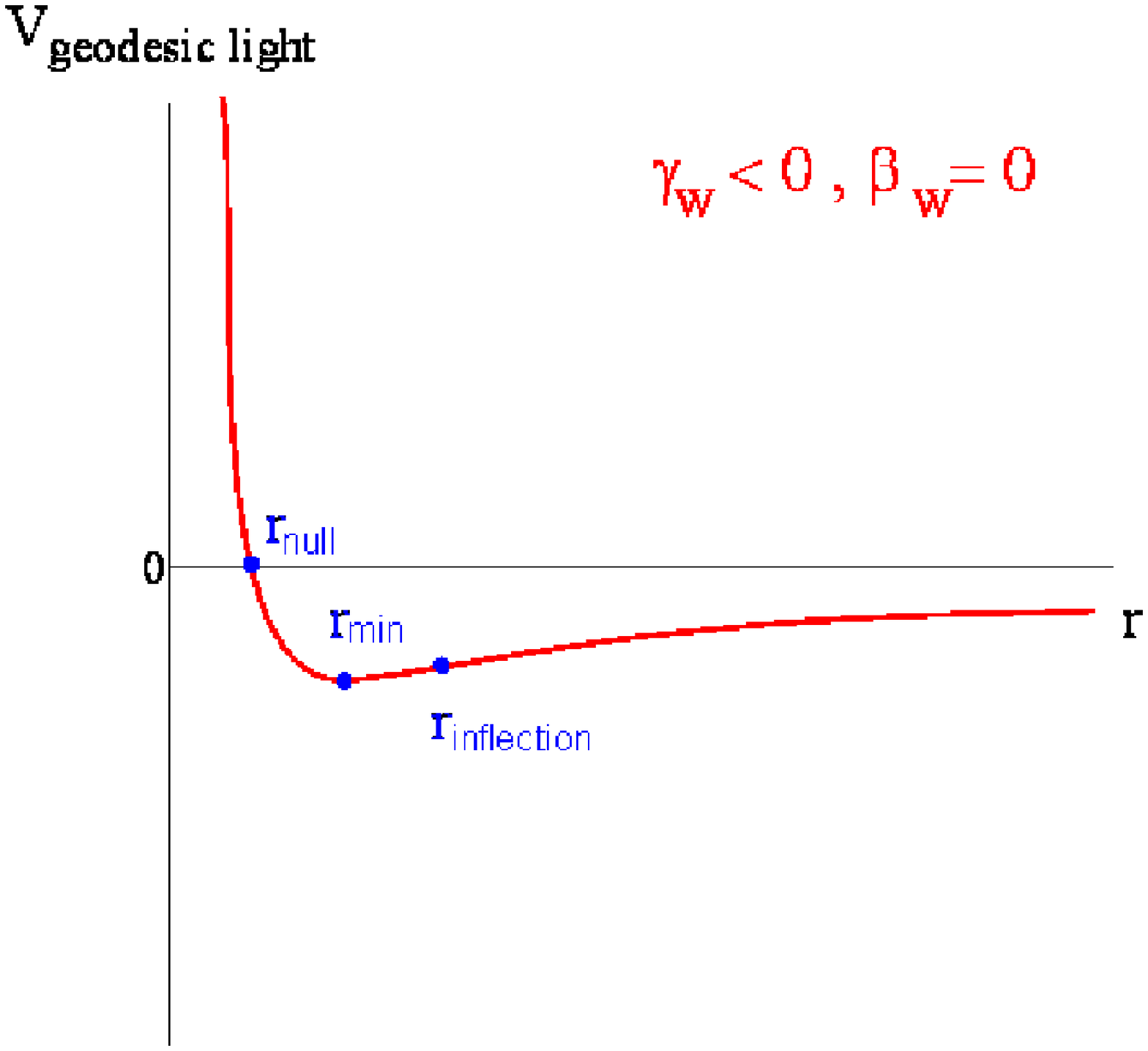';file-properties "XNPEU";}%
}

\FRAME{fhFU}{2.2701in}{2.5573in}{0pt}{\Qcb{The strong field regime.
Asymptotic photon trajectories for a positive value of the Weyl parameter $%
\gamma _{W}$ and the deflector located at the origin of the coordinate
system. All the orbits are hyperbolic and divergent. In this plot, $%
r_{0}=\frac{1}{2\ \gamma _{W}}$, $r_{0}=\ \frac{1}{\gamma _{W}}$, $r_{0}=%
\frac{3}{2\ \gamma _{W}}$, $r_{0}=\frac{2}{\ \gamma _{W}}$, and $r_{0}=\frac{%
5}{2\ \gamma _{W}}$ respectively for the dash-dot curve, solid line, dash
curve, bold curve and dotted line.}}{\Qlb{bis_photon_trajectories_weyl}}{%
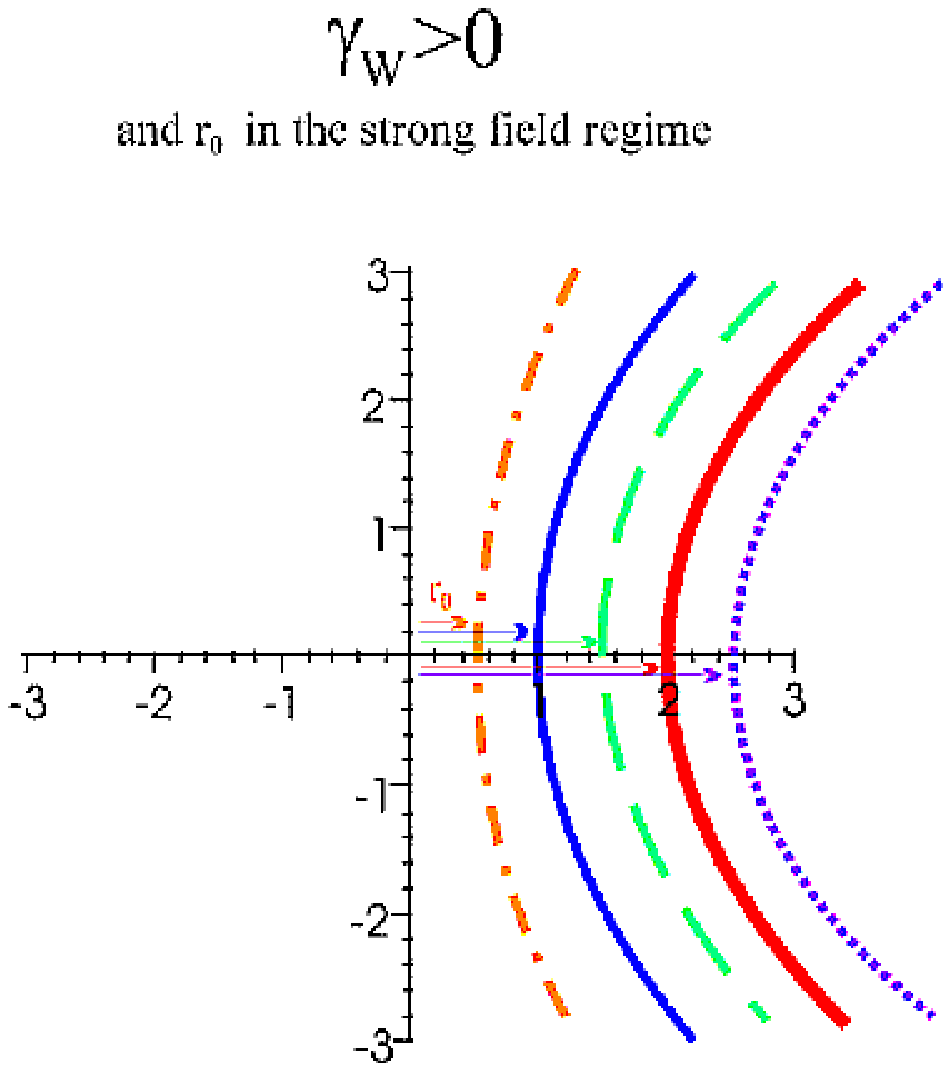}{\special{language "Scientific Word";type
"GRAPHIC";maintain-aspect-ratio TRUE;display "ICON";valid_file "F";width
2.2701in;height 2.5573in;depth 0pt;original-width 0pt;original-height
0pt;cropleft "0";croptop "1";cropright "1";cropbottom "0";filename
'bis_photon_trajectories_weyl.eps';file-properties "XNPEU";}%
}

\FRAME{fhFU}{2.2701in}{2.5867in}{0pt}{\Qcb{The strong field regime.
Asymptotic photon trajectories for a negative value of the Weyl parameter $%
\gamma _{W}$ and the deflector located at the origin of the coordinate
system. With respect to the characteristic distances $r_{null}\equiv
-2/\gamma _{W}$ and $r_{\min }\equiv -1/\gamma _{W}$, the orbits are
convergent and elliptic for $r_{0}>r_{null}$ ( in particular, circular for $%
r_{0}=r_{\min }$), parabolic for $r_{0}=r_{null}$ and hyperbolic for $%
r_{0}<r_{null}$. Hence, light deflection is possible only when $r_{0}\leq
r_{null}$. Note that all the orbits have the same semi-lattice rectum
(position located at an angle $\pi /2$ from the closest approach distance
position) value of $2/\left| \gamma _{W}\right| $ . In this plot, $%
r_{0}=\frac{-1}{2\ \gamma _{W}}$, $r_{0}=\ \frac{-1}{\gamma _{W}}$, $r_{0}=%
\frac{-3}{2\ \gamma _{W}}$, $r_{0}=\frac{-2}{\ \gamma _{W}}$, and $r_{0}=%
\frac{-5}{2\ \gamma _{W}}$ respectively for the dash-dot curve, solid line,
dash curve, bold curve and dotted line.}}{\Qlb{photon_trajectories_weyl}}{%
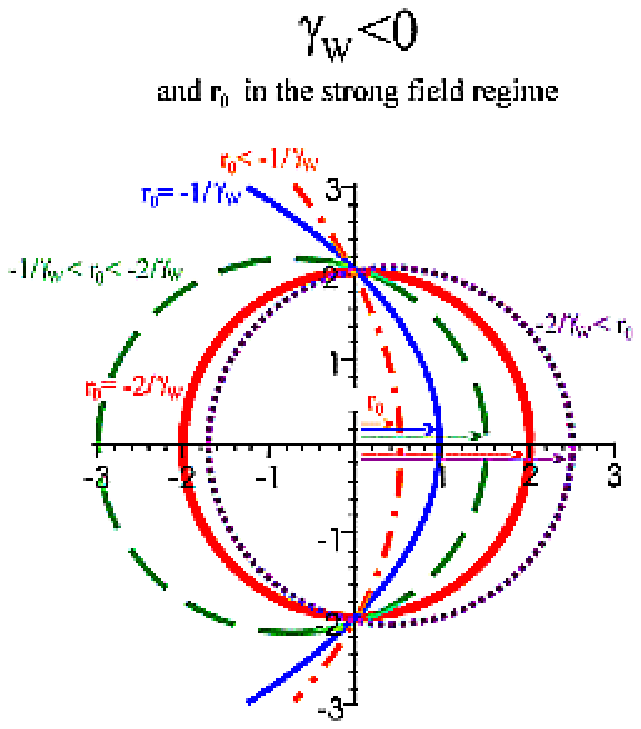}{\special{language "Scientific Word";type
"GRAPHIC";maintain-aspect-ratio TRUE;display "ICON";valid_file "F";width
2.2701in;height 2.5867in;depth 0pt;original-width 0pt;original-height
0pt;cropleft "0";croptop "1";cropright "1";cropbottom "0";filename
'photon_trajectories_weyl.eps';file-properties "XNPEU";}%
}
\pagebreak 

\FRAME{fhFU}{2.4146in}{3.1989in}{0pt}{\Qcb{The intermediate regime. The
semi-lattice rectum, located at $r_{\min }=-2/\gamma _{W}$ for any orbit in
the strong field limit, is an accumulation point. In the weak field limit,
light deflection for a point mass is similar to the general relativistic
predictions: light converges on short distances when it passes close to the
gravitational mass, while it converges on farther distances when it has a
larger closest approach distance from the lens. We may guess the
intermediate regime.}}{\Qlb{accumulation_point_weyl}}{%
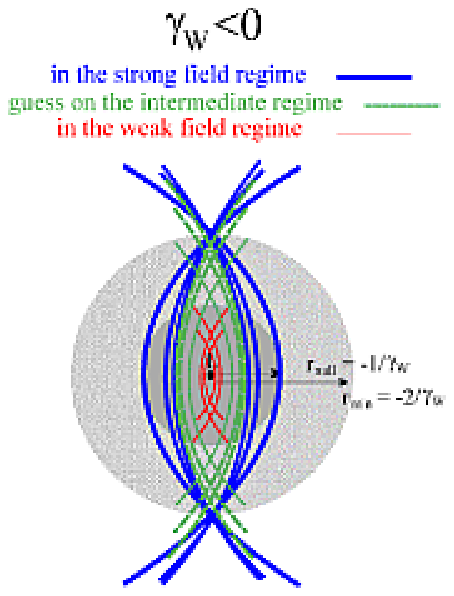}{\special{language "Scientific Word";type
"GRAPHIC";maintain-aspect-ratio TRUE;display "ICON";valid_file "F";width
2.4146in;height 3.1989in;depth 0pt;original-width 0pt;original-height
0pt;cropleft "0";croptop "1";cropright "1";cropbottom "0";filename
'accumulation_point_weyl.eps';file-properties "XNPEU";}%
}

\FRAME{fhFU}{2.2649in}{2.8988in}{0pt}{\Qcb{In the strict strong field regime
($\beta _{W}=0$), there exists a particular O-L-S alignment configuration in
which the observer and the source aligned with the lens are located on the
semi-lattice rectum points. There would then be an infinite number of photon
trajectories coming from the source to the observer.}}{\Qlb{%
particular_ols_weyl}}{particular_ols_weyl.eps}{\special{language "Scientific
Word";type "GRAPHIC";maintain-aspect-ratio TRUE;display "ICON";valid_file
"F";width 2.2649in;height 2.8988in;depth 0pt;original-width
532pt;original-height 682.5625pt;cropleft "0";croptop "1";cropright
"1";cropbottom "0";filename
'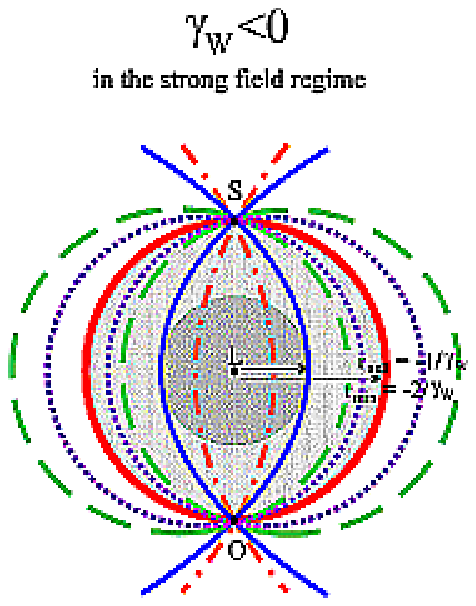';file-properties "XNPEU";}%
}
\pagebreak

\FRAME{fhFU}{5.124in}{6.4351in}{0pt}{\Qcb{In the strong field regime, for a
light source, (S$_{1}$) located inside the sphere of radius $r_{null}$, the
closest approach distance of any photon originating from the source is
necessarily smaller than $r_{null}$. This means that all these photons have
unbound orbits: hyperbolas centered on L.\newline
When the source (S$_{2}$) is on the sphere of radius $r_{null}$, the closest
approach distance of its photons can be either smaller or equal to $r_{null}$%
, which means that all the orbits are again unbound: hyperbolas or a
parabola.\newline
Whereas for a source (S$_{3}$) located in between the two spheres of radius $%
r_{null}$ and $r_{\min }$, there are additionally photons originating from
this source that have a closest approach distance to the lens larger than $%
r_{null}$. Accordingly, those photons are captured on elliptic orbits. The
particular case of a source (S$_{4}$) situated on the sphere of radius $%
r_{\min }$ allows all four types of orbits: hyperbolic, parabolic, circular
and elliptic. The photon trajectories will be again hyperbolic, parabolic or
elliptic for a source (S$_{5}$) outside the larger sphere of radius $r_{\min
}$. }}{\Qlb{visibility_ols_weyl_12345}}{visibility_ols_weyl_12345.eps}{%
\special{language "Scientific Word";type "GRAPHIC";maintain-aspect-ratio
TRUE;display "ICON";valid_file "F";width 5.124in;height 6.4351in;depth
0pt;original-width 543pt;original-height 682.5625pt;cropleft "0";croptop
"1";cropright "1";cropbottom "0";filename
'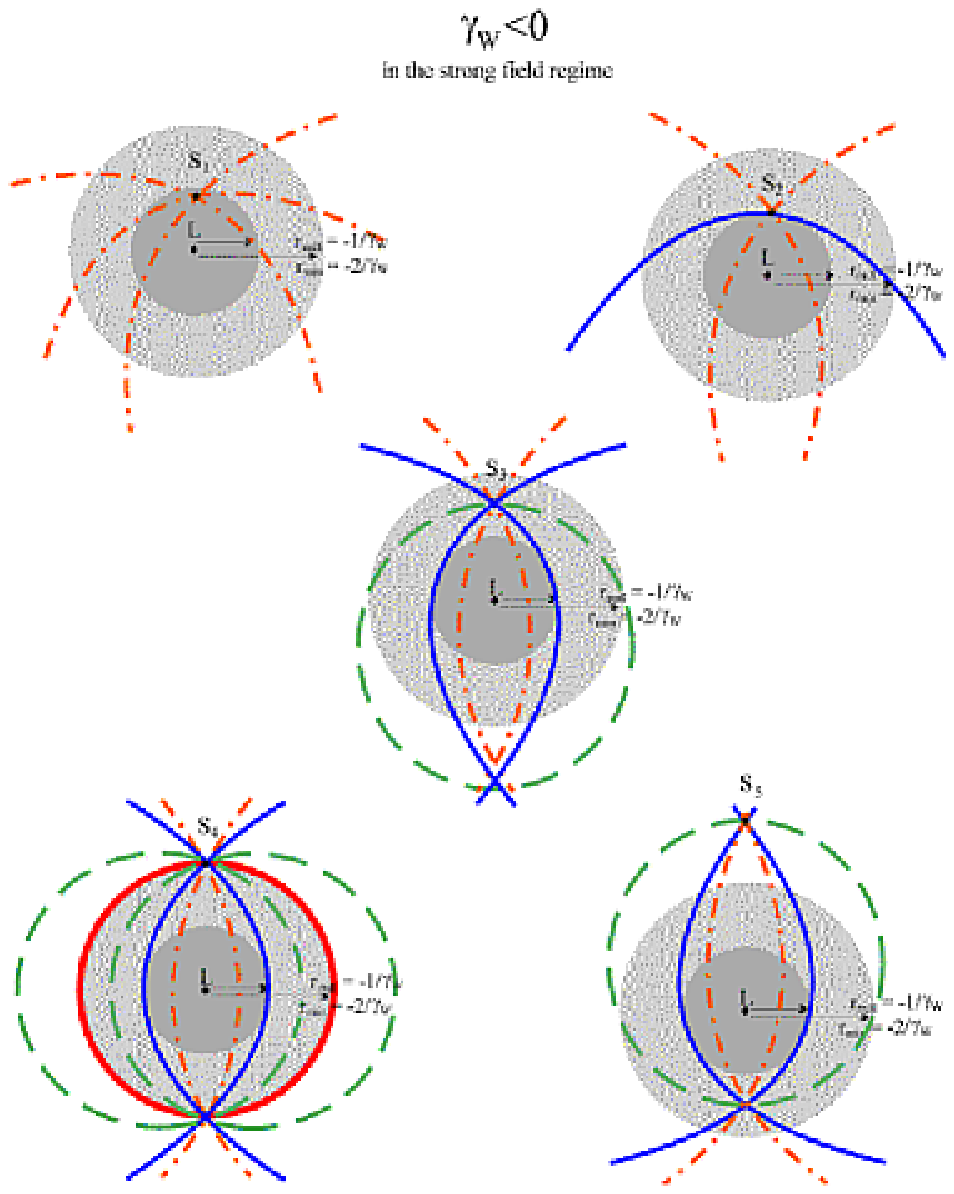';file-properties "XNPEU";}%
}
\pagebreak

\clearpage

\ack{%
The research work presented here was carried out in the FYMA
Institute at the University of Louvain la Neuve, during a Ph.D.thesis
financed by the I.I.S.N research assistantship. We are thankful to Professor
J-M. G\'{e}rard (FYMA, UCL) for pertinent advice and carefully proofreading
the manuscript.}


\bigskip

\begin{thebibliography}{}
\bibitem[EdPa1998]{Edery 1998 weyl causal structure}  A. Edery and M. B.
Paranjape. Causal structure of vacuum solutions to conformal (Weyl)
gravity. General Relativity and Gravitation, 31, 1031 (1999)
astro-ph/9808345 v2.

\bibitem[FrTs1981]{Fradkin 1981 weyl asymp free}  E. S. Fradkin and A. A.
Tseytlin. \textit{Renormalizable asymptotically free quantum theory of
gravity}. Physics Letters B, 104, 377-381 (1981).

\bibitem[FrTs1982]{Fradkin 1982 Weyl renormalizable}  E. S. Fradkin and A.
A. Tseytlin. \textit{Renormalizable asymptotically free quantum theory of
gravity.} Nuclear Physics B, 201, 5, 469-491 (1982).

\bibitem[JuTo1978]{Julve 1978 weyl asymp free}  J. Julve and M. Tonin. 
\textit{Quantum gravity with higher derivative terms}. Nuovo Cimento B, 46,
1, 137-152 (1978)

\bibitem[Ma1994a]{Mannheim 1994 open questions}  P. D. Mannheim. \textit{%
Open Questions in Classical Gravity}. Foundations of Physics, 24, 4, 487-511
(1994).

\bibitem[Ma1994b]{Mannheim 1994 microlensing}  P. D. Mannheim. \textit{%
Microlensing, Newton-Einstein gravity, and conformal gravity.} UCONN-94-10,
astro-ph/9412007 (1994).

\bibitem[Ma1995]{Mannheim 1995 Age of Universe}  P. D. Mannheim. \textit{%
Conformal Cosmology and the Age of the Universe}. UCONN 95-08 (1995).

\bibitem[Ma1996]{Mannheim 1997 Global Gravity}  P. D. Mannheim. \textit{%
Local and Global Gravity}. Foundations of Physics, 26, 12, 1683-1709 (1996).

\bibitem[Ma2002]{Mannheim 2002 conformal factor}  P. D. Mannheim. Private
Communication. University of Connecticut, August 2002.

\bibitem[MaKa1989]{Mannheim 1989 Exact Solution}  P. D. Mannheim and D.
Kazanas. \textit{Exact Vacuum Solution to Conformal Weyl Gravity and
Galactic Rotation Curves}. Astrophysical Journal, 342, 635-638 (1989).

\bibitem[MaKa1991]{Mannheim 1991 Equations of motion}  P. D. Mannheim and D.
Kazanas. \textit{General Structure of the Gravitational Equations of Motion
in Conformal Weyl Gravity}. Astrophysical Journal, Supplement Series, 76,
431-453 (1991).

\bibitem[Pi1997]{Pireaux 1997 memoire}  S. Pireaux. \textit{Etude de
Solutions Particuli\`{e}res d'une Th\'{e}orie Invariante Conforme de la
Gravitation}. Graduate thesis, Universit\'{e} catholique de Louvain (UCL),
June 1997.

\bibitem[Pi2002]{Pireaux 2002 thesis}  S. Pireaux. \textit{Light deflection
experiments as a test of relativistic theories of gravitation}. Ph.D.
thesis, Universit\'{e} catholique de Louvain (UCL), September 2002.

\bibitem[Pi2003]{Pireaux 2003 constraints on Weyl parameter}  S. Pireaux. 
\textit{Light deflection in Weyl gravity: constraints on the linear
parameter.} Submited to Classical and Quantum Gravity.

\bibitem[St1977]{Stelle 1977 renormalization}  K. S. Stelle. \textit{%
Renormalization of higher order derivative quantum gravity}. Physical Review
D, 16, 4, 953-969 (1977).

\bibitem[Wi2001]{Will 2001 summary of tests}  C. M. Will. The confrontation
between General Relativity and experiments. Living Reviews, 4
(2001). \TEXTsymbol{\backslash}\TEXTsymbol{\backslash}%
www.livingreviews.org/Articles/Volume4/2001-4will
\end{thebibliography}
\end{document}